\documentclass[twocolumn,english,aps,prb,showpacs]{revtex4}
\usepackage{ae}
\usepackage{aecompl}
\usepackage[T1]{fontenc}
\usepackage{graphicx}

\makeatletter

\usepackage{babel}
\makeatother
\begin{document}

\title{Stability of Jahn-Teller distortion in LaMnO$_{3}$ under pressure:
direct study by x-ray absorption }

\author{Aline Y. Ramos}

\email{aline.ramos@grenoble.cnrs.fr}

\affiliation{Institut Néel, UPR 2940 - CNRS, 25 av. des Martyrs, BP166, 38042
Grenoble, France}

\affiliation{Laboratório Nacional de Luz Síncrotron - LNLS, P.O. Box 6192, 13084-971,
Campinas, São Paulo, Brazil}

\author{Hélio C. N. Tolentino}

\affiliation{Institut Néel, UPR 2940 - CNRS, 25 av. des Martyrs, BP166, 38042
Grenoble, France}

\affiliation{Laboratório Nacional de Luz Síncrotron - LNLS, P.O. Box 6192, 13084-971,
Campinas, São Paulo, Brazil}

\author{Narcizo M. Souza-Neto}

\affiliation{Laboratório Nacional de Luz Síncrotron - LNLS, P.O. Box 6192, 13084-971,
Campinas, São Paulo, Brazil}

\affiliation{Dept. de Física dos Materiais e Mecânica, DFMT-IF-USP, São Paulo,
SP, Brazil}

\author{Jean-Paul Itié}

\affiliation{Synchrotron SOLEIL, L'Orme des Merisiers, Saint-Aubin, BP 48, 91192
Gif-sur-Yvette Cedex, France}

\author{Liliana Morales}

\affiliation{Centro Atómico Bariloche and Instituto Balseiro, Comisión Nacional
de Energía Atómica and Universidad Nacional de Cuyo, 8400 S.C. de
Bariloche, Argentine.}

\author{Alberto Caneiro}

\affiliation{Centro Atómico Bariloche and Instituto Balseiro, Comisión Nacional
de Energía Atómica and Universidad Nacional de Cuyo, 8400 S.C. de
Bariloche, Argentine.}

\date{\today{}}

\begin{abstract}
The local environment of manganese atoms in LaMnO$_{3}$ under pressure
up to 15.3~GPa has been studied by x-~ray absorption spectroscopy.
For pressures below 8~GPa, no change is detected within the MnO$_{6}$~octahedra.
Above this pressure a continuous reduction of the long Mn-O distance
takes place, however the octahedral distortion persists over the whole
pressure range. At 15.3 GPa the average Jahn-Teller splitting of the
distances is reduced by about one third, indicating that a total removal
of the local Jahn-Teller distortion would occur only for pressures
around 30~GPa, where metallization is reported to take place. A hysteresis
in the long distance reduction is observed down to ambient pressure,
suggesting the coexistence of MnO$_{6}$ distorted and undistorted
units. 
\end{abstract}


\maketitle
The physics underlining the remarkable properties of the manganite
LaMnO$_{3}$ and its doped alloys is rich and complex. The actual
path followed by a given system towards ferromagnetism and/or metallization,
through chemical substitution, thermal treatment or pressure involves
modifications of a delicate balance between delocalization and carriers
trapping not yet completely understood. In the ground state LaMnO$_{3}$
is an antiferromagnetic semiconductor crystallizing in an orthorhombic
variant of the cubic perovskite structure space group $Pbnm$. The
MnO$_{6}$ octahedra are distorted due to the Jahn-Teller ($JT)$
effect of the Mn$^{3+}$($t_{2g}^{3}e_{g}^{1}$) and the $Mn-O$ distances
are split into 4 short bonds and 2 long bonds. In the basal \textit{ab}
plane long and short $Mn-O$ bonds alternates. The apical and basal
short bonds have different length, however this additional distortion
is not resolved by local probes such as real space high resolution
diffraction and x-ray absorption spectroscopy. The local radial distribution
actually seen by these probes corresponds then to the $JT$ splitting,
with 4 oxygensat short distances $(Mn-O){}_{s}$$\approx$ 1.94~$\textrm{\AA}$
and 2 oxygens at the long distance $(Mn-O){}_{l}\approx$~2.15~$\textrm{\AA}$.
LaMnO$_{3}$ undergoes a transition at $T^{*}$ $\approx$~710-750
K from the $JT$ distorted orthorhombic phase O to a high temperature
nearly cubic O' phase \cite{Rodriguez-Carvajal-PRB98}. The transition
is accompanied by abrupt changes in the electrical resistivity and
Weiss constant \cite{Zhou-PRB99}. The cell distortion is nearly removed
and the orbital ordering disappears in the O' phase, but the~$JT$
distortion of MnO$_{6}$ octahedra persists at the local scale \cite{Araya-Rodriguez-JMMM01,Sanchez-PRL03,Souza-PRB04}.
The transition then happens as an order-disorder transition, in agreement
with the thermodynamic calculations \cite{Millis-PRB96}. More recently
Qiu and co-workers \cite{Qiu-PRL05} reported on neutron powder diffraction
measurements showing that the $JT$ distortion of MnO$_{6}$ octahedra
is maintained also in the high temperature rhomboedral phase ($T$
$\geq$~1010 K) and suggested the presence of ordered clusters with
strong antiferrodistorsive coupling. 

New insights for the role of the $JT$ distortion can be obtained
by the exploration of its pressure dependence. In LaMnO$_{3}$, the
application of an external hydrostatic pressure produces a reduction
of the lattice distortions and an enhancement of the carrier mobility
\cite{Loa-PRL01,Pinsard-PRB01}. The $Mn-O-Mn$ angle - tilt angle
among two adjacent octahedra - is reported to decrease monotonically.
The short bond distances $(Mn-O){}_{s}$~are weakly shortened with
increasing pressure, the largest effect being a shortening of the
long distance $(Mn-O){}_{l}$. Besides, resistivity measurements \cite{Loa-PRL01}
reveal that the system keeps its insulating nature at room temperature
up to 32 GPa where it undergoes an insulator- metal transition. From
the extrapolation of x-ray diffraction data, it has been inferred
that local $JT$ distortion completely vanishes around 18 GPa \cite{Loa-PRL01}.
However \textit{ab initio} calculations using soft pseudo-potentials
recently predicted the conservation of local octahedral distortion
well above this value\cite{Trimarchi-PRB05}. The authors predict
the occurrence of a structural phase transition around 15~GPa leading
to a situation with a mixture of polytypes of \textit{~}\textit{\emph{antiferromagnetic}}
order\cite{Mizokawa-PRB99}. Based on mean field theoretical calculations,
Yamasaki and coworkers \cite{Yamasaki-PRL06} also claimed that pressure
induced metal to insulator transition in LaMnO$_{3}$ is not of Mott
Hubbard type. They show that, according to calculations combining
local density approximations and mean field theories, both on site
repulsion and Jahn-Teller distortion are necessary for LaMnO$_{3}$~to
be insulating below 32~GPa. The issue of local distortion in LaMnO$_{3}$~at
high pressure is then not fully addressed. 

Although the use of x-~ray absorption spectroscopy (XAS), has been
conclusive to elucidate many critical points of the local structure
of LaMnO$_{3}$ and its doped alloys\cite{Tyson-PRB96,Subias-PRB97,Booth-PRB98,Shibata-PRB03},
no pressure dependence of the local order has been reported yet by
XAS measurements, neither in the XANES (x-ray absorption near edge
structure) nor in the EXAFS (extended x-ray absorption fine structure)
range. This is principally due to inherent experimental difficulties
in the collection of the XAS high pressures data at the manganese
\emph{K} edge, arising both from the low transmission of the diamond
cells and from the strong additional absorption due to the La \emph{L}
edges. These difficulties were partially overcome here by using perforated-diamond
cell \cite{Dadashev-RSI01,Itié-JPC05} and taking advantage of the
high stability of a dispersive XAS setup. This setup is based on a
bent crystal monochromator that opens up the energy band pass necessary
for the spectroscopy and focuses it to the sample position. The dispersing
band-pass is collected by a linear detector giving rise to a full
spectrum at once. There is no moving optical element during experiments
and an extremely good accuracy in the energy scale can be achieved
\cite{TOLENTINO-JAC88}. Above 8~GPa the long distance $(Mn-O){}_{l}$
is continuously reduced, however an octahedral distortion persists
over the whole pressure range. A hysteresis in the long distance reduction
is observed down to ambient pressure, suggesting the coexistence of
MnO$_{6}$ distorted and undistorted units.

The pressure dependent x-ray absorption measurements at the Mn \emph{K}
edge were performed at XAS dispersive beamline \cite{Tolentino-PS05}
of the \textit{LNLS} (Laboratório Nacional de Luz Síncrotron, Campinas,
Brazil). A polycrystalline powder sample of ~LaMnO$_{3}$~ was synthesized
by the liquid-~mix method previously described \cite{Prado-JSSChem99}.
The as-made sample was annealed at \emph{T}=~1000~$^{\circ}$C under
oxygen partial pressure \emph{p}($O_{2}$) during 24 h and then quenched
at liquid nitrogen temperature. The \emph{T} and \emph{p}($O_{2}$)
values were chosen in order to give~ an LaMnO$_{3}$~oxygen stoichiometric
compound, according to high temperature thermodynamic measurements
\cite{Morales-JSSChem03}. The lattice parameters obtained from the
Rietveld refinement agree with those of literature \cite{Prado-JMMM99}.
A fine grained powder sample was loaded in a 100~$\mu m$ diameter
hole of an iconel gasket mounted on perforated diamond as support
to the 500 $\mu m$ thick diamond anvils\cite{Itié-JPC05}. Silicone
oil was used as pressure transmitive medium. Quasi - hydrostatic pressures
up to 15.3~GPa were applied and controlled using ruby sphere with
a precision of about 0.5~GPa. For each pressure, the cell was then
realigned at the optical focus - of around 150~$\mu m$ - and the
spectra were collected in the transmission geometry. The beam path
was set under vacuum, to reduce air absorption and beam harmonic contamination.
To ensure the beam harmonic purity an additional gold coated mirror
was added just behind the anvil cell and set to a grazing angle of
8 mrad. The energy resolution was about 1.5~eV, with energy calibration
stable within 0.05~eV during the experiments. 

\begin{figure}
\includegraphics[scale=0.9]{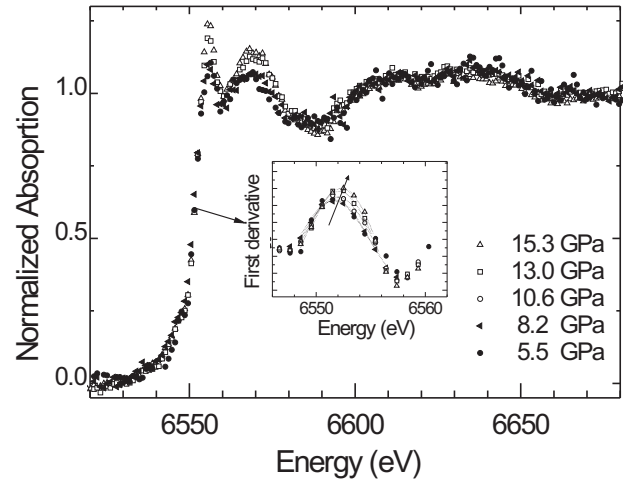}

\caption{\label{fig1-pressure}Mn \emph{K} edge XANES spectra for LaMnO$_{3}$
at increasing pressures. Up to 8~GPa, the XANES are unchanged. Above
8~GPa, the structures just beyond the edge (6650 - 6680~eV) are
slighly enhanced, whereas the absorption threshold is continuously
shifted towards higher energies. The inset shows the spectra derivatives
for a best observation of this last effect.}
\end{figure}

The XANES spectra collected at various pressures exhibit slight modifications
(Fig.\ref{fig1-pressure}), indicating that the local atomic manganese
environment is not drastically changed under external hydrostatic
pressure. For pressures up to 8~GPa, no change at all is detected
within the MnO$_{6}$~octahedra. In this pressure range a reduction
of the cell volume has been observed by X-ray and neutron diffraction
measurements\cite{Loa-PRL01,Pinsard-PRB01}. A continuous decrease
of the $Mn-O$ bond length from ambient pressure to 8GPa is also reported.
Even if they are  sharply contrasted, there may not be contradiction between
the diffraction and the XAS results. On the first hand XRD provides
the static periodicity of the structure averaged over a large spatial
domain. On the other hand, as the characteristic time in the photoabsorption
process is smaller than that the time corresponding to the thermal
motion of the atomic bonds, XAS probes the instantaneous short range
structure around the absorbing atoms. A good example of complementarity of these approaches is given by LaMnO$_{3}$
on crossing the Jahn-Teller transition temperature. Diffraction methods
show that the three bond converge into a single bond length on crossing
$T^{*}$\cite{Rodriguez-Carvajal-PRB98}, whereas XAS gives no change
of all in the three bonds lengths\cite{Araya-Rodriguez-JMMM01,Sanchez-PRL03,Souza-PRB04}.
The dynamic nature of the Jahn-Teller transition has been deduced
from the confrontation of both experimental evidences. By analogy
with the temperature dependent measurements, we may assume here that
when an external pressure above 8GPa is applied the main effect should
be then the rearrangement among the octahedra, with possibly the formation
of domains of disordered distorsions, while the local instantaneous
octahedron keeps almost unchanged. Such rearrangment would result
in a reduction of the coherence length of the dynamical spatial fluctuations,
and yield, in diffraction measurements, to the derivation of smaller
average static values for the $Mn-O$ bonds.

Above 8~GPa, we observe a continuous shift of the absorption threshold
towards higher energies, (Fig.\ref{fig1-pressure} and inset), along
with the enhancement of the structures just above the edge (6650 -
6680~eV). As the manganese formal valence (Mn$^{3+}$) keeps unchanged
with pressure, the edge shift ($\delta$E) expresses here modifications
in the repulsive nearest neighbors potential arising from change in
$Mn-O$ bonds in the coordination shell\cite{Natoli-84}. Besides,
as shorter bond lengths correspond to higher edge energies, the edge
position is determined by the long distance $(Mn-O){}_{l}\approx$~2.15~$\textrm{\AA}$.
$\delta$E is then here essentially related to the specific reduction
in this bond. At the same time, the structures close to the edge are
enhanced. It should be pointed out that, due to the selection rules
in x-~ray absorption spectroscopy, the \emph{K} edge transition originates
from the core $1s$ state to the projected $np$ (mainly $4p$) unoccupied
states. The enhancement in the structures close to the edge accounts
for a change in the hybridization concomitant with the increasing
overlap of the wavefunctions when the hydrostatic pressure is applied.
This indicates a reduction of the local distortion of the Mn sites,
leading to a reduction of the $e_{g}$ splitting and a partial delocalization
of the $e_{g}$~electrons. Our results then agree with the intuitive
idea, also confirmed by the x-ray diffraction measurements\cite{Pinsard-PRB01,Loa-PRL01},
that the short bonds $(Mn-O){}_{s}$ should be less reduced than the
long $(Mn-O){}_{l}$ ones by the application of an external pressure.

\begin{figure}[h]
\includegraphics[scale=0.87]{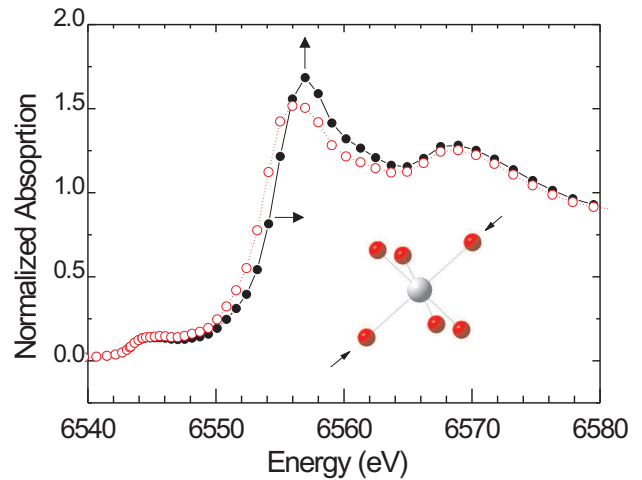}

\caption{\label{fig2-simul}XANES simulations. Open symbols : LaMnO$_{3}$~structure.
Plain symbols : LaMnO$_{3}$ variant where the $(Mn-O){}_{l}$ bonds
are reduced ($\approx$ 0.1~$\textrm{\AA}$). The edge is shifted
($\approx$ 0.9~\emph{eV}) and the structures above the edge are
enhanced.}
\end{figure}

\begin{figure}
\includegraphics[clip,scale=0.68]{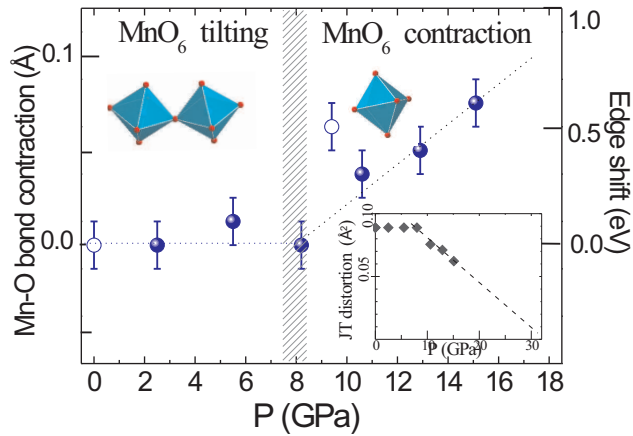}

\caption{\label{fig3-bond contraction}Relative changes in $Mn-O$~distance
in the pressure range 0 to 15.3 GPa. The plain circles are obtained
by increasing the pressure and the open circles correspond to the
release. The curve in the inset gives an estimation of the $JT$ distortion
defined by $(\sigma{}_{JT}$)$^{2}$ =($\sqrt{1/6\cdot{\displaystyle \sum\mid R{}_{i}-{}R\mid}}$
)$^{2}$ and predicts a suppression of the distortion above 30~GPa}
\end{figure}

\textit{Ab initio} XANES calculations\cite{Ankudinov-PRB98} of LaMnO$_{3}$-~based
structures with progressive reduction of the $(Mn-O){}_{l}$ bond
reproduce well the experimental features (Fig.\ref{fig2-simul}).
We should report that in simulations where long and short bonds are
reduced in a same proportion, the positive edge shift is correctly
reproduced but the structures at the edge are not enhanced. The evolution
of the XANES features reflects then a continuous reduction of the
average $JT$ distortion from 8~GPa up to 15.3~GPa. The evolution
of the edge energy as a function of the applied pressure is given
in figure \ref{fig3-bond contraction}. For small shifts the relationship
between the reduction $\delta$R of a bond distance and the associated
edge shift $\delta$E and is almost linear\cite{Natoli-84}. An experimental
calibration obtained for manganite systems \cite{SouzaNeto-PRB04}
gives $\frac{\delta E}{\delta R}\approx$ 10 ~\emph{eV}$/\textrm{\AA}$.
We also obtain the similar calibration from our XANES simulations
($\frac{\delta E}{\delta R}\approx$ 9~\emph{eV}$/\textrm{\AA}$,
Fig.\ref{fig2-simul}). 

The edge shift of 0.6 \emph{eV} from 8 to 15.3 GPa corresponds then
to a reduction in the long bond $(Mn-O){}_{l}$ by about 0.06~$\textrm{\AA}$$\pm$~0.02~$\textrm{\AA}$
over this range. Even if the short bonds $(Mn-O){}_{s}$ were not
reduced at all under pressure, the total suppression of the $JT$
splitting would result in an energy shift at the edge of the order
of 2~\emph{eV}, which is not observed.~Up to 15~GPa, the coherent
$JT$ distortion parameter $(\sigma{}_{JT}){}^{2}$ - defined as $\sigma{}_{JT}$
=$\sqrt{1/6\cdot{\displaystyle \sum\mid R{}_{i}-R\mid}}$ - decreases
only by one third (Fig.\ref{fig3-bond contraction}, inset). Local
$JT$ distortions are then present, even if the crystallographic structure
suggests otherwise. Extrapolation of the data shows that the $JT$
splitting would vanish only around 30~GPa, \emph{i.e.} for applied
pressures where the system is reported to undergo an insulator - metal
transition \cite{Loa-PRL01}. 

When the pressure is released down to ambient, a hysteresis is observed.
At 9~GPa all features of the XANES spectrum are similar to those
of the spectra at 13~GPa. Such hysteresis suggests the occurrence
of a mixture of phases with close compositions and related structures\cite{Grunbaum-JSSChem04}.
The hysteresis results in this case from the non-negligible elastic
strain energy of coherent or semi-coherent interfaces that must be
taken into account to describe the total Gibbs energy of the system.~As
the free motion of each MnO$_{6}$ octahedron is limited by the oxygen
atoms shared with the adjacent units, electron phonon coupling may
involve several coupled Mn atoms \cite{Sanchez-PRB06}. We should
remind that XAS informs on the average Mn environment. The present
data may then be examined considering that application of the pressure
above 8~GPa, would induce the progressive formation of MnO$_{6}$
undistorted units coupled to distorted ones. The coexistence, at high
pressures, of these larger polarons with the small MnO$_{6}$ $JT$
distorted octahedra may account for the hysteresis behavior. 

In summary, we studied the modifications in the average local distortion
around the manganese atoms induced by application of high pressures
in the range 0 to 15.3 GPa, from the modifications of x-ray absorption
spectra in the near edge range. The MnO$_{6}$ octahedra keep unchanged
by application of hydrostatic pressures up to 8~GPa, whereas above
this value the long $(Mn-O)_{l}$~distance is continuouly reduced.
The Jahn-~Teller bond splitting persists over the whole pressure
range. The total quenching of this splitting is expected to take place
only at pressures above 30~GPa, indicating that local $JT$ distortion
and insulator - to - metal transition should be closely related. A
hysteresis in the XANES features when pressure is released suggests
the coexistence of MnO$_{6}$ distorted and undistorted units.

\begin{acknowledgments}
This work is partially supported by LNLS/ABTLuS/MCT, CNPq and CNPq-CNRS
agreement. \bibliographystyle{apsrev}

\end{acknowledgments}

\end{document}